\documentclass[prl,aps, twocolumn,floatfix,groupedaddress,superscriptaddress]{revtex4-1}
\usepackage[english]{babel}
\usepackage{graphicx}
\usepackage{amsmath,bbm,amssymb,multirow,times,bm}

\begin{document}

\title{Quantum support vector machine for big data classification}
\author{Patrick Rebentrost}
\email{rebentr@mit.edu}
\affiliation{Research Laboratory of Electronics,
Massachusetts Institute of Technology, Cambridge, MA 02139}
\author{Masoud Mohseni}
\affiliation{Google Research, Venice, CA 90291}
\author{Seth Lloyd}
\email{slloyd@mit.edu}
\affiliation{Department of Mechanical Engineering, Massachusetts Institute of Technology, Cambridge, MA 02139}

\begin{abstract}
Supervised machine learning is the classification of new data based on
already classified training examples.  
In this work, we show that the support vector machine, an optimized 
binary classifier, can be implemented on a quantum computer, with
complexity logarithmic in the size of the vectors and the number of training 
examples. In cases when classical sampling algorithms require polynomial time, 
an exponential speed-up is  obtained. At the core of this quantum big data algorithm 
is a non-sparse matrix exponentiation technique for efficiently performing 
a  matrix inversion of the training data inner-product (kernel) matrix. 
\end{abstract}

\maketitle

Machine learning algorithms can be categorized along a spectrum of supervised and
unsupervised learning
\cite{Mackay2003,Alpaydin2004,Bishop2007,Murphy2012}.  In strictly unsupervised
learning, the task is to find structure such as clusters in unlabeled data.
Supervised learning involves a
training set of already classified data, from which inferences are
made to classify new data. In both cases, recent ``big data" applications exhibit
a growing number of features and input data.   A support vector machine
(SVM) is a supervised machine learning algorithm that classifies
vectors in a feature space into one of two sets, given training data
from the sets \cite{Cortes1995}.  It operates by constructing
the optimal hyperplane dividing the two sets, either in the original
feature space or a higher-dimensional kernel space.  The SVM can be formulated as a quadratic programming problem
\cite{Boyd2004}, which can be solved in time proportional to $O(\log(\epsilon^{-1}) {\rm
  poly} (N,M))$, with $N$ the dimension of the feature space, $M$ the
number of training vectors, and $\epsilon$ the accuracy. 
In a quantum setting, binary classification
was discussed in terms of Grover search in \cite{Anguita2003} and 
using the adiabatic algorithm in \cite{Neven2008,Neven2009,Pudenz2011,Denchev2012}. 
Quantum learning was also discussed in \cite{Sasaki2002,Servedio2004}.

In this paper, we show that a quantum support vector machine can
be implemented with $O(\log N M)$ run time in both training and classification
stages.  
The performance in $N$ arises due to a fast quantum
evaluation of inner products, 
discussed in a general machine learning
context by us in \cite{Lloyd2013}.  
For the performance in
$M$, we re-express the SVM as an approximate least-squares problem
\cite{Suykens1999} that allows for a quantum solution with the matrix
inversion algorithm \cite{Harrow2009,Wiebe2012}.  We
employ a technique for the exponentiation of
non-sparse matrices recently developed in \cite{Lloyd2013_2}.  
This allows us to reveal efficiently in quantum form the largest eigenvalues and 
corresponding eigenvectors of the training data overlap (kernel) and covariance matrices. 
We thus efficiently perform a  
low-rank approximation of these matrices (principal component analysis, PCA). 
PCA is a common task arising here and in other machine learning algorithms
\cite{Muller2001,Hoegaerts2004,Hofmann2008}. 
The error dependence in the training stage is
 $O({\rm poly}(\epsilon_K^{-1}, \epsilon^{-1}))$, 
 where $\epsilon_K$ is the smallest eigenvalue considered and $\epsilon$ is the accuracy.
In cases when a low-rank approximation is appropriate, our quantum SVM operates on the full training set in logarithmic runtime.

\emph{Support vector machine.$-$} 
The task for the SVM is to classify a vector into one of two classes, given $M$ training data
points of the form $\{ (\vec x_j, y_j): \vec x_j \in \mathbbm{R}^N,
y_j = \pm 1\}_{j=1 \ldots M}$, where $y_j = 1$ or $-1$ depending on
the class to which $\vec x_j$ belongs.  
For the classification, the SVM finds a maximum-margin
hyperplane with normal vector $\vec w$ that divides the two classes. 
The margin is given by two parallel hyperplanes 
that are separated by the maximum possible distance $2/|\vec w|$ 
with no data points inside the margin. 
Formally, these hyperplanes are constructed
so that $\vec w \cdot \vec x_j + b \geq 1$ for $\vec x_j$ in the $+1$
class and that $\vec w \cdot \vec x_j + b \leq -1$ for $\vec x_j$ in
the $-1$ class, where $b/|\vec w|$ is the offset of the hyperplane. 
Thus, in the primal formulation,
finding the optimal hyperplane consists of minimizing $|\vec
w|^2/2$ subject to the inequality constraints $y_j(\vec w \cdot \vec
x_j + b) \geq 1$ for all $j$. 
The dual formulation \cite{Boyd2004} is maximizing over 
the Karush-Kuhn-Tucker multipliers
 $\vec{\alpha}=(\alpha_1,\cdots,\alpha_M)^T$ the function:
\begin{equation} \label{eqDual}
L(\vec{\alpha})= \sum_{j=1}^M y_j \alpha_j - \frac{1}{2} \sum_{j,k=1}^M \alpha_j   K_{jk} \alpha_k
,
\end{equation}
subject to the constraints $\sum_{j=1}^M \alpha_j =0$ and $y_j \alpha_j
\geq 0$. 
The hyperplane parameters are recovered from  $\vec w
= \sum_{j=1}^M \alpha_j \vec x_j$ and $b = y_j - \vec w \cdot
\vec x_j$ (for those $j$ where $\alpha_j\neq0$).
Only a few of the $\alpha_j$ are non-zero: these are the ones
corresponding to the $\vec x_j$ that lie on the two hyperplanes -- the
support vectors. 
We have introduced the kernel matrix, a central quantity for
supervised machine learning problems \cite{Muller2001,Hofmann2008},
$K_{jk}=k(\vec x_j,\vec x_k)=\vec x_j \cdot \vec x_k$, defining the
kernel function $k(x,x')$. More complicated non-linear kernels and
soft margins will be studied below.  Solving the dual form involves
evaluating the $M(M -1)/2$ dot products $\vec x_j \cdot \vec x_k$ in
the kernel matrix, and then finding the optimal $\alpha_j$ values by
quadratic programming, which takes $O(M^3)$ in the non-sparse case
\cite{ [{The exponent $3$ can be improved considerably: }]
  Coppersmith1990}.  As each dot product takes time $O(N)$ to 
evaluate, the classical support vector algorithm takes time
$O(\log(1/\epsilon) M^2 (N+M))$ with accuracy $\epsilon$.  The result is a binary classifier for new data $\vec x$:
\begin{eqnarray}\label{eqSVMClassifier}
y(\vec x) = {\rm sign} \left ( \sum_{j=1}^M \alpha_j k(\vec{x}_j, \vec{x} ) + b \right).
\end{eqnarray}

\emph{Quantum machine learning with the kernel matrix.$-$} 
In the quantum setting, assume that oracles for the
training data that return quantum vectors $|\vec{x}_j\rangle = 1/|\vec
x_j| \sum_{k=1}^N (\vec x_j)_k |k\rangle$, the norms $|\vec{x}_j|$,
and the labels $y_j$ are given. 
The quantum machine learning performance is relative to these oracles and can be considered
a lower bound for the true complexity 
\cite{Aaronson2009}. 
One way of efficiently constructing these states 
is via quantum RAM, which uses $O(MN)$ hardware resources but only $O(\log MN)$ operations to access them, see \cite{Giovannetti2008,Lloyd2013}. 
Using the inner product evaluation of  \cite{Lloyd2013} to prepare the kernel matrix, 
we can achieve a run time for the SVM
of $O(\log(1/\epsilon)M^3+M^2\log N/\epsilon)$. Classically, the inner
product evaluation is $O(\epsilon^{-2}{\rm poly}(N))$ by sampling when the components of 
the $\vec x_j$ are distributed unevenly, for example 
when a Fourier transform is part of the post-processing step \cite{Aaronson2009}.

The kernel matrix plays a crucial role in the dual formulation
Eq. (\ref{eqDual}) and the least-squares reformulation discussed in
the next section. At this point we can  discuss an efficient quantum method for
direct preparation and exponentiation of the normalized kernel matrix
$\hat K=K/{\rm tr}K$.  
For the  preparation, first
call the training data oracles with the state $1/\sqrt{M}\sum_{i=1}^M
|i\rangle$. This prepares in quantum parallel the state $|\chi\rangle
=1/\sqrt{N_\chi}\sum_{i=1}^M |\vec{x}_i| |i\rangle |\vec x_i\rangle$,
with $N_\chi=\sum_{i=1}^M |\vec{x}_i|^2$, in $O(\log NM)$ run time.  
If we discard the training set register, we
obtain the desired kernel matrix as a quantum density matrix. This can
be seen from the partial trace
$ {\rm tr}_2 \{ |\chi\rangle \langle \chi |\} 
= \frac{1}{N_\chi} \sum_{i,j=1}^M \langle \vec{x}_j |  \vec{x}_i\rangle |\vec{x}_i||\vec{x}_j| |i\rangle \langle j| =\frac{K}{{\rm tr} K}$.
See the appendix A for an independent estimation of the trace of $K$.

For quantum mechanically computing a matrix inverse such as $\hat
K^{-1}$ one needs to be able to enact $e^{-i \hat K
  \Delta t}$ efficiently. However, the kernel matrix $\hat{K}$ is not sparse for
the application of the techniques in
\cite{Childs2010,Berry2007}. For the exponentiation of non-sparse symmetric or Hermitian
 matrices a strategy was developed by us in
\cite{Lloyd2013_2}.  We adapt it to the present problem. Adopting a
density matrix description to extend the space of possible
transformations gives, for some quantum state $\rho$,
$e^{-i \hat K \Delta t}\  \rho \ e^{i\hat K \Delta t} = e^{-i\mathcal{L}_{\hat K} \Delta t} (\rho)$,
The super-operator notation $\mathcal{L}_{K}(\rho) =[K,\rho]$
was defined.  Applying the
algorithm of \cite{Lloyd2013_2} obtains:
\begin{eqnarray}\label{eqTimeEvolutionPractice}
e^{-i\mathcal{L}_{\hat K} \Delta t} (\rho) 
&\approx& {\rm tr}_1 \{ e^{-i S \Delta t} \hat{K} \otimes \rho e^{iS \Delta t } \}  \\
&=& \rho- i \Delta t [\hat{K}, \rho]  + O(\Delta t^2). \nonumber
\end{eqnarray}
Here, $S=\sum_{m,n=1}^M |m\rangle \langle n| \otimes |n\rangle\langle
m|$ is the swap matrix of dimension $M^2 \times M^2$.
Eq. (\ref{eqTimeEvolutionPractice}) is the operation that is
implemented on the quantum computer performing the machine
learning. For the time slice $\Delta t$, it consists of the
preparation of an environment state $\hat{K}$ (see above) and the
application of the global swap operator to the combined
system/environment state followed by discarding the environmental
degrees of freedom.  This shows that enacting $e^{-i \hat K
  \Delta t}$ is possible with error $O(\Delta t^2)$.  The efficient
preparation and exponentiation of the training data kernel matrix, which
appears in many machine
learning problems \cite{Muller2001,Hofmann2008}, potentially enables a
wide range of supervised quantum machine learning algorithms.
We now discuss a complete quantum big data algorithm.

\emph{Quantum least-squares support vector machine.$-$} A key idea of this work is to employ the least-squares reformulation
of the support vector machine developed in \cite{Suykens1999} that
circumvents the quadratic programming and obtains the parameters from
the solution of a linear equation system.  The central simplification
is to introduce slack variables $e_j$ and replace the inequality
constraints with equality constraints (using $y_j^2=1$):
\begin{equation}\label{eqEqualityConstraints}
y_j(\vec w \cdot \vec x_j + b) \geq 1 \to (\vec w \cdot \vec x_j  + b) =  y_j -  y_j e_j.
\end{equation}
In addition to the constraints, the implied Lagrange function contains
a penalty term $\gamma/2 \sum_{j=1}^M e_j^2$, where user-specified
$\gamma$ determines the relative weight of training error and SVM
objective. Taking partial derivatives of the Lagrange function and
eliminating the variables $\vec u$ and $e_j$ leads to a
least-squares approximation of the problem:
\begin{equation}\label{eqLeastSquareSVM}
F
\left ( \begin{array}{c}
b \\
\vec{\alpha} 
\end{array}
\right )
\equiv 
\left ( \begin{array}{cc}
0 & \vec{1}^T \\
\vec{1} & K + \gamma^{-1}  \mathbbm{1}
\end{array}
\right ) 
\left ( \begin{array}{c}
b \\
\vec{\alpha} 
\end{array}
\right )
=
\left ( \begin{array}{c}
0 \\
\vec{y} 
\end{array}
\right ).
\end{equation}
Here, $K_{ij} = \vec{x}_i^T \cdot \vec{x}_j$ is again the symmetric
kernel matrix, $\vec{y}=(y_1,\cdots,y_M)^T$, and
$\vec{1}=(1,\cdots,1)^T$.  The matrix $F$ is $(M+1)\times( M+1)$
dimensional.  The additional row and column with the $\vec 1$ arise
because of a non-zero offset $b$.  
The $\alpha_j$ take on the role as distances from the optimal margin 
and usually are not sparse.
The SVM
parameters are determined schematically by $ \left (b,
\ \vec{\alpha}^T \right)^T = F^{-1} \left( 0, \vec{y}^T \right )^T$.
As with the quadratic programming formulation, the classical complexity of the
least-squares support vector machine is $O(M^3)$
\cite{Coppersmith1990}.

For the quantum support vector machine, the task is to generate a
quantum state $|b,\vec{\alpha} \rangle$ describing the hyperplane with
the matrix inversion algorithm \cite{Harrow2009} and then classify a
state $|\vec x\rangle$.  
We solve the normalized $\hat F |b,\vec{\alpha} \rangle = |\vec y\rangle$, where $\hat F= F/{\rm tr}F $ with
$||F|| \leq 1$.
The classifier will be determined by the success probability of a swap test between 
$|b,\vec{\alpha} \rangle$ and $|\vec x\rangle$.
For application of the quantum matrix inversion
algorithm  $\hat F$ needs to be exponentiated efficiently. 
The matrix $\hat F$ is schematically separated as
$\hat F=(J+K +\gamma^{-1} \mathbbm{1})/{\rm trF}$, with $J=
\left ( \begin{array}{cc}
0 & \vec{1}^T \\
\vec{1} &  0
\end{array}
\right )$, and the Lie product formula allows for $e^{-i \hat F \Delta t}=e^{-i \Delta t  \mathbbm{1} /{\rm trF}}
  e^{-iJ\Delta t /{\rm trF} }e^{-i K \Delta t/{\rm trF} }+ O(\Delta t^2)$.
  The matrix $J$ is straightforward \cite{Childs2010} (``star" graph). 
The two nonzero eigenvalues of $J$
are $\lambda_{\pm}^{\rm star}=\pm \sqrt{M} $ and the corresponding
eigenstates are $| {\lambda_{\pm}^{\rm star}}
\rangle=\frac{1}{\sqrt{2}} \left( |0\rangle \pm \frac{1}{\sqrt{M}}
\sum_{k=1}^M |k\rangle \right)$.  The matrix $\gamma^{-1}
\mathbbm{1}$ is trivial.  For  $K/{\rm tr}K$, proceed according to Eq. 
(\ref{eqTimeEvolutionPractice}) by rescaling time by a factor $\frac{{\rm tr} K}{{\rm tr} F}=O(1)$ appropriately.  
This $e^{-i\hat F \Delta t}$ is employed conditionally in  phase estimation.

The right hand side $|\vec y\rangle$  can be formally expanded  into eigenstates $|u_j\rangle$ of
$\hat F$ with corresponding eigenvalues $\lambda_j$,
$|\tilde{y}\rangle = \sum_{j=1}^{M+1} \langle u_j | \tilde{y} \rangle
|u_j\rangle$.  With a register for storing an approximation of the
eigenvalues (initialized to $ |0\rangle$),  phase estimation generates
 a state which is close to the ideal state storing the respective eigenvalue:
\begin{equation} \label{eqStateInverted}
|\tilde{y}\rangle |0\rangle \to \sum_{j=1}^{M+1} \langle u_j |  \tilde{y} \rangle |u_j\rangle |\lambda_j\rangle \to   \sum_{j=1}^{M+1} \frac{\langle u_j |  \tilde{y} \rangle}{\lambda_j} |u_j\rangle.
\end{equation}
The second step inverts the eigenvalue and is obtained as in \cite{Harrow2009} by performing a controlled rotation and uncomputing the eigenvalue register. 
In the basis of training set labels, the expansion coefficients of the new state are
the desired support vector machine parameters: ($C= b^2+\sum_{k=1}^M \alpha_k^2$)
\begin{equation}\label{eqQuantumHyperplane}
| b,\vec{\alpha} \rangle = \frac{1}{\sqrt{C}} \left( b |0\rangle + \sum_{k=1}^M \alpha_k |k\rangle \right ).
\end{equation}

\emph{Classification.$-$} We have now trained the quantum SVM
 and would like to classify a query
state $| \vec x \rangle$.  From the state $|b, \vec \alpha\rangle$ in
Eq. (\ref{eqQuantumHyperplane}), 
construct by calling the training-data oracle:
\begin{equation}
|\tilde{u}\rangle=\frac{1}{\sqrt{N_{\tilde{u}}}} \left( b|0\rangle|0\rangle + \sum_{k=1}^M \alpha_k |\vec x_k| |k\rangle |\vec x_k\rangle \right),
\end{equation}
with $N_{\tilde{u}}= b^2+\sum_{k=1}^M \alpha_k^2|\vec x_k|^2$. 
In addition, construct the query state:
\begin{equation}
| \tilde{x} \rangle =\frac{1}{\sqrt{N_{ \tilde{x}} }} \left( |0\rangle |0\rangle + \sum_{k=1}^M |\vec x| |k\rangle |\vec x\rangle \right).
\end{equation}
with $N_{ \tilde{x}} = M|\vec x|^2+1$.  For the classification, we
perform a swap test. Using an ancilla, construct the state $|\psi
\rangle=\frac{1}{\sqrt{2}}( |0\rangle |\tilde{u} \rangle + |1\rangle
|\tilde{x}\rangle)$ and measure the ancilla in the state $|\phi
\rangle=\frac{1}{\sqrt{2}}( |0\rangle - |1\rangle)$.  The measurement
has the success probability $P= | \langle \psi |\phi\rangle |^2 =
\frac{1}{2} (1 - \langle \tilde{u} | \tilde{x} \rangle)$.  The inner
product is given by $\langle \tilde{u}| \tilde{x} \rangle =
\frac{1}{\sqrt{N_{ \tilde{x}} N_{\tilde{u}}}} \left( b+\sum_{k=1}^M
\alpha_k |\vec x_k| |\vec x|\langle \vec x_k| \vec x\rangle
\right)$, which is $O(1)$ in the usual case when the $\alpha$ are not sparse.
$P$ can be obtained to accuracy $\epsilon$ by iterating $O(P(1-P)/\epsilon^2)$ times. 
If $P < 1/2$ we classify  $|\vec x\rangle$ as $+1$, otherwise $-1$.

\emph{Kernel matrix approximation and error analysis.$-$} We now show that quantum matrix inversion essentially
performs a kernel matrix principal component analysis and give a run
time/error analysis of the quantum algorithm.  The matrix under
consideration, $\hat F=F/{\rm tr} F$, contains the kernel matrix $\hat
K_\gamma=K_\gamma/ {\rm tr} K_\gamma$ and an additional row and column
due to the offset parameter $b$.  In case the offset is negligible,
the problem reduces to matrix inversion of the kernel matrix $\hat
K_\gamma$ only.  For any finite $\gamma$, $\hat K_\gamma$ is positive definite, 
and thus invertible.  
The positive eigenvalues of $\hat F$ are
dominated by the eigenvalues of $\hat K_\gamma$.  In addition, $\hat
F$ has one additional negative eigenvalue which is involved in
determining the offset parameter $b$.  
The maximum absolute eigenvalue of $\hat F$
is no greater than $1$ and the minimum absolute eigenvalue is $\leq O(1/M)$.  The
minimum eigenvalue arises e.g. from the possibility of having a training
example that has (almost) zero overlap with the other training
examples. Because of the normalization the eigenvalue will be $O(1/M)$
and as a result the condition number $\kappa$ (largest eigenvalue divided by smallest eigenvalue)
is  $O(M)$ in this case. To resolve such an eigenvalue 
would require exponential runtime
\cite{Harrow2009}.  We define a constant $\epsilon_K$
such that only the eigenvalues in the interval $\epsilon_K \leq |
\lambda_j| \leq 1$ are taken into account, essentially defining an
effective condition number $\kappa_{\rm eff} = 1/\epsilon_K$.  Then,
the filtering procedure described in \cite{Harrow2009} is employed in
the phase estimation using this $\kappa_{\rm eff}$.  An ancilla register is attached to the quantum state and appropriately defined filtering functions
discard eigenvalues below $\epsilon_K$ 
when multiplying the inverse $1/\lambda_j$ for each eigenstate in
Eq. (\ref{eqStateInverted}). The desired outcome is
obtained by post-selecting the ancilla register.

The legitimacy of this eigenvalue filtering can be rationalized by 
a principal component analysis (PCA) argument.   
Define the $N\times M$ (standardized) data matrix $X=\left( \vec x_1, \cdots, \vec x_M \right)$.
The $M\times M$ kernel matrix is given by $K= X^T X$. The $N\times N$
covariance matrix is given by $\Sigma=X X^T=\sum_{m=1}^M \vec x_m \vec x^T_m$. 
Often data sets are effectively described by a few unknown factors (principal components), 
which admit a low-rank approximation for $\Sigma$.
This amounts to finding the eigenvectors $\vec v_i$ of $\Sigma$ 
with the largest eigenvalues $\lambda_i$.  The matrices $XX^T$ and $X^T X$ 
have the same non-zero eigenvalues.
Keeping the large eigenvalues and corresponding eigenvectors of the kernel
matrix thus retains the principal components of the covariance matrix, i.e. the most important features of the data. See Appendices B and C for further discussion of the low-rank approximation.
Regarding the cutoff $\epsilon_K$, 
 $O(1)$ eigenvalues of $K/{\rm tr} K$ exist for example  in the case  
of well-separated clusters with $O(M)$ vectors in them. 
A simple artificial example is $K=\mathbbm{1}_{2\times 2} \otimes(\vec 1 \vec 1^T)_{M/2\times M/2}$.
Note that finding the principal components of the kernel matrix  is performed in quantum parallel
by phase estimation and the filtering procedure.  

We continue with a discussion of the run time of the quantum
algorithm. The interval $\Delta t$ can be written as $ \Delta t =
t_0/T$, where $T$ is the number of time steps in the phase estimation
and $ t_0$  is the total evolution time determining the error of the phase
estimation \cite{Harrow2009}.  The swap matrix used in
Eq. (\ref{eqTimeEvolutionPractice}) is $1$-sparse and $e^{-i S \Delta
  t}$ is efficiently simulable in negligible time $\tilde O(\log
(M) \Delta t) $ \cite{Berry2007}.  The $\tilde O$ notation suppresses
more slowly growing factors, such as a $\log^{\ast} M$ factor
\cite{Harrow2009,Berry2007}.  For the phase estimation, the propagator
$e^{-i\mathcal{L}_{\hat F} \Delta t} $ is enacted with error
$O(\Delta t^2 ||\hat F||^2 )$, see Eq.
(\ref{eqTimeEvolutionPractice}). With the spectral norm for a matrix
$A$, $||A||=\max_{|\vec v|=1} |A\vec v|$, we have $||\hat
F||=O(1)$. Taking powers of this propagator, $e^{-i\mathcal{L}_{\hat
    F} \tau \Delta t} $ for $\tau=0,\cdots, T-1$, leads to an error of
maximally $\epsilon=O(T \Delta t^2 )=O( t_0^2 /T)$. Thus, the run time
is $T=O( t_0^2 /\epsilon)$.  Taking into account the preparation of
the kernel matrix in $O(\log MN)$, the run time is thus $ O( t_0^2
\epsilon^{-1} \log M N )$. The relative error of $\lambda^{-1}$ by phase
estimation is given by $O(1/(t_0 \lambda))\leq O(1/(t_0 \epsilon_K))$ for
$\lambda\geq \epsilon_K$ \cite{Harrow2009}. If $t_0$ is taken $O(\kappa_{\rm
  eff}/\epsilon)=O(1/(\epsilon_K \epsilon))$ this error is
$O(\epsilon)$.  The run time is thus $\tilde O(\epsilon_K^{-2}\epsilon^{-3} \log M N)$. 
Repeating the algorithm for $O(\kappa_{\rm
  eff})$ times to achieve a constant success probability of the
post-selection step obtains a final run time of $ O( \kappa_{\rm
  eff}^3 \epsilon^{-3} \log M N) $.  To summarize, we find a quantum
support vector machine that scales as $O(\log M N)$, which implies a
quantum advantage in situations where classically $O({\rm poly}M)$ 
training examples and $O(N)$ samples for the inner product  are required.

\emph{Nonlinear support vector machines.$-$} One of the most powerful uses of support vector machines is to perform
nonlinear classification \cite{Cortes1995}.  Perform a nonlinear
mapping $\vec \phi(\vec x_j)$ into a higher-dimensional vector space.
Thus, the kernel function becomes a nonlinear function in $\vec x$:
\begin{equation}
k(\vec x_j, \vec x_k) = \vec \phi(\vec x_j) \cdot \vec \phi(\vec x_k).
\end{equation}
For example, $k(\vec x_j, \vec x_k) = (\vec x_j \cdot \vec x_k)^d$.
Now perform the SVM classification in the
higher-dimensional space.  The separating hyperplanes in the
higher-dimensional space now correspond to separating nonlinear
surfaces in the original space.

The ability of quantum computers to manipulate high-dimensional
vectors affords a natural quantum algorithm for polynomial kernel
machines.  Simply map each vector $|\vec x_j\rangle$ into the
$d$-times tensor product $|\phi(\vec x_j)\rangle \equiv |\vec
x_j\rangle \otimes \ldots \otimes |\vec x_j\rangle$ and use the
feature that $\langle \phi(\vec x_j) | \phi(\vec x_k) \rangle =
\langle \vec x_j | \vec x_k\rangle^d$.  Arbitrary polynomial kernels
can be constructed using this trick.  The optimization using a
nonlinear, polynomial kernel in the original space now becomes a
linear hyperplane optimization in the $d$-times tensor product space.
Considering only the complexity in the vector space dimension, the
nonlinear $d$-level polynomial quantum kernel algorithm to accuracy
$\epsilon$ then runs in time $O(d \log N/\epsilon)$.  Note that,
in contrast to classical kernel machines, the exponential quantum
advantage in evaluating inner products allows quantum kernel machines
to perform the kernel evaluation directly in the higher dimensional
space.

\emph{Conclusion.$-$} 
In this work, we have shown that an important classifier in machine
learning, the support vector machine, can be implemented quantum
mechanically 
with algorithmic complexity logarithmic in feature size
and the number of training data, thus providing one example of a
quantum ``big data" algorithm.  A least-squares
formulation of the support vector machine allows the use of
phase estimation and the quantum matrix inversion algorithm. 
The speed of the
quantum algorithm is maximized when the training data kernel matrix is
dominated by a relatively small number of principal components.  
We note that there exist several heuristic sampling algorithms for the SVM \cite{Shalev2007} and, more 
generally, for finding eigenvalues/vectors of low-rank matrices \cite{Liberty2007,Drineas2006_2}. 
Information-theoretic arguments show that classically finding a low-rank matrix approximation 
is lower-bounded by $\Omega(M)$ in the absence of prior knowledge \cite{Yossef2003}, suggesting a similar lower bound for the least-squares SVM.
Aside from the speed-up, another timely benefit of quantum machine learning is data privacy \cite{Lloyd2013}.
The
quantum algorithm never requires the explicit $O(MN)$ representation of all
the features of each of the training examples, but generates the
necessary data structure, the kernel matrix of inner products, in
quantum parallel. 
Once the kernel matrix is generated, the individual
features of the training data are fully hidden from the user.
In summary,  the quantum support vector machine is an
efficient implementation of an important machine learning
algorithm. It also provides advantages in terms of data privacy and
could be used as a component in a larger quantum neural network.
Recently, quantum machine learning was discussed in \cite{Wiebe2014,Paparo2014}.

This work was supported by DARPA, NSF, ENI, Google-NASA Quantum Artificial Intelligence Laboratory, and Jeffrey Epstein.
The authors acknowledge helpful discussions with Scott Aaronson and Nan Ding.

\bibliography{QML}

\section{Appendix A: Estimating the trace of the kernel matrix}
\label{appendixTraceK}
The trace of the kernel matrix can be efficiently evaluated, similar
to \cite{Lloyd2013}. Generate the Hamiltonian $H_{\rm tr} =
\sum_{j=1}^M |\vec x_j| |j\rangle \langle j|\otimes \sigma_x$ from the
quantum access to the norms due to the training-data oracle.  Applying
$e^{-i H_{\rm tr} t}$ to the state $|\psi \rangle=1/\sqrt{M}
\sum_{j=1}^M |j\rangle | 0\rangle$ results in
$|\psi(t)\rangle=1/\sqrt{M} \sum_{j=1}^M\left( \cos( |\vec x_j| t)
|j\rangle |0\rangle-i \sin(|\vec x_j| t) |j\rangle |1\rangle
\right)$. Choose $t$ such that $|\vec x_j|t \ll 1$, for all $j$, and
measure the ancilla in the $|1\rangle$ state. This succeeds with
probability $1/M \sum_{j=1}^M |\vec x_j|^2 t^2$, which allows the
estimation of the trace of $K$, $\sum_{j=1}^M |\vec x_j|^2$.

\section{Appendix B: Low-rank approximation}
To investigate the kernel matrix low-rank approximation, 
first set $b=0$ and $K=0$. From $\gamma^{-1} I \vec \alpha = \vec y$ and $\vec w = \sum _m \alpha_m \vec x_m$, the normal vector is then simply  
$\vec w =\gamma \sum_m  y_m  \vec x_m$ with consequently a 
sub-optimally small margin $2/|\vec w|$. Now assume a rank-one approximation for the kernel matrix,
$K\approx\lambda_1\vec{u}_1 \vec{u}_1^T$. Note that the vector $\vec{u}_1$ 
is related to the first principal component $\vec v_1$ (first eigenvector of the covariance matrix $\Sigma=XX^T$) by 
$\vec{u}_1 = X^T \vec v_1 / \sqrt{\lambda_1} =1/\sqrt{\lambda_1} (\vec x_1^T \vec v_1,\cdots, \vec x_M^T \vec v_1)^T$ \cite{Ding2004}, i.e.
its elements are the inner products of the training vectors with  $\vec v_1$.
This low-rank approximation for $K$ gives $(K+\gamma^{-1} I)^{-1}=\gamma( I-c \vec{u}_1 \vec{u}_1^T)$, with  $c=\gamma \lambda_1/(1+\gamma\lambda_1)$, using the Sherman-Morrison formula. 
This leads to $\vec w = \gamma \sum_m ( y_m - c' \vec u_{1,m})  \vec x_m$, with 
$c'=c \vec u_1^T \vec y$, which corrects \textit{each} $y_m$ by taking into 
account the respective projection of the training 
example on the first principal component. Thus it is shown that in the low-rank 
approximation all training examples nevertheless
 are contributing to the solution. That is in contrast to stochastic 
gradient approaches \cite{Shalev2007}, which only sample a small subset of the training examples.

\section{Appendix C: Error of quantum low-rank approximation}

Our quantum speed-up holds in the case when the data is in a low-rank situation, i.e. the kernel matrix has a few $O(1)$ eigenvalues and 
many $O(1/M)$ eigenvalues, all of them initially unknown. The quantum algorithm only takes into account the $O(1)$ eigenvalues, which incurres an error E.  This error is 
given by the norm of the difference between
 the low-rank matrix $K$ and its quantum approximation $K_q$, i.e. $E=||K - K_q||_F$, using the 
Frobenius/Hilbert-Schmidt norm. This error is given by $E=\sqrt{\sum_{\lambda_i = O(1/M)}\lambda_i^2}$, where $\lambda_i$ are the eigenvalues of $K$.
Since by assumption we have $O(M)$ small eigenvalues, this error is $E = O(1/\sqrt{M})$.

\end{document}